\documentclass[12pt]{iopart}

\usepackage{graphicx}
\usepackage{iopams}

\begin{document}

\title[Magneto-optical trapping forces for atoms and molecules]{Magneto-optical trapping forces for atoms and molecules with complex level structures}

\author{M. R. Tarbutt}

\address{Centre for Cold Matter, Blackett Laboratory, Imperial College London, Prince Consort Road, London SW7 2AZ, United Kingdom}

\ead{m.tarbutt@imperial.ac.uk}

\begin{abstract}
Laser cooling and magneto-optical trapping of molecules typically involves multiple transitions driven by several laser frequencies. We analyze how magneto-optical trapping forces depend on the angular momenta, $F_l$ and $F_u$, and the g-factors, $g_l$ and $g_u$, of the lower and upper states. When $F_l > F_u$ the polarizations must be reversed relative to cases where $F_u \ge F_l$. The correct choice of circular polarization depends on the sign of $g_{u}$ but not on the sign of $g_{l}$. If $g_{u}$ is zero there is no trapping force, and the trapping force is very weak whenever $g_u$ is small compared to $g_l$, which it usually is when the cooling transition is the $^{2}\Sigma$ to $^{2}\Pi_{1/2}$ transition of a molecule. For some molecules, mixing of the excited $^{2}\Pi_{1/2}$ state with a nearby $^{2}\Sigma$ excited state can greatly increase $g_u$, leading to stronger trapping forces. A strong trapping force can also be produced by rapidly and synchronously reversing both the magnetic field and the laser polarizations. We simulate a recent experiment on magneto-optical trapping of SrF molecules, and suggest that an alternative choice of laser beam polarizations will strengthen the trapping force.
\end{abstract}

\maketitle

\section{Introduction}

The magneto-optical trap (MOT) \cite{Raab(1)87} is a crucial tool for a huge number of experiments and measuring devices based on ultracold atoms. In a 1D MOT, atoms interact with a pair of counter-propagating, circularly polarized laser beams, each slightly red detuned from an atomic transition, and a magnetic field gradient sets up a spatially-dependent Zeeman splitting of the transition. The red-detuning ensures that atoms are Doppler cooled, and the combination of the red detuning, circular polarization, and Zeeman shift, sets up an imbalance in the radiation pressure from the two beams, ensuring that atoms are always pushed towards the zero of magnetic field. A standard 3D MOT repeats this arrangement in all three dimensions, using three orthogonal pairs of counter-propagating beams and a quadrupole magnetic field. In this way, large numbers of atoms can be captured from a hot gas, trapped in a small volume, typically less than 1\,mm radius, and cooled to a low temperature, typically below 1\,mK. It is important to avoid optical pumping into a dark state, where for a given polarization the transition matrix element to all relevant excited states is zero. In almost all MOTs, the main cooling cycle is a transition between lower and upper levels with angular momenta $F_{l}$ and $F_{u}=F_{l}+1$. For alkali atoms for example, the cooling transition is typically the $^{2}S_{1/2} -\, ^{2}P_{3/2}$ D$_2$ transition. Since there are no dark Zeeman sub-levels in this case, the atoms are always in the cooling cycle. Often, the upper state can decay to some other lower state, either a metastable excited state, or another hyperfine level of the ground state. Repump lasers are then used to keep the atoms in the cooling cycle. These atomic MOTs have been extensively studied and are well understood.

Atomic MOTs where the cooling transition has $F_{u} \le F_{l}$ are also sometimes used \cite{Prentiss(1)88, Shang(1)94, Flemming(1)97, Nasyrov(1)01, Atutov(1)01, Tiwari(1)08}. Examples are MOTs working on the $^{2}S_{1/2}-^{2}P_{1/2}$ D$_1$ transition of alkali atoms. The confining forces in these MOTs tend to be weaker than in a normal atomic MOT because an atom cannot scatter photons continuously from any one beam without being optically pumped into a state that is dark to that beam, but bright to the opposing beam (see section \ref{Sec:AngMomCases}). In these cases, the orthogonal beams can be important in turning dark states into bright ones to set up a net restoring force. The changing magnetic field orientation and laser polarization seen by the atoms as they move through the MOT can also play that role, and mixing of closely-spaced hyperfine levels by the magnetic field can contribute to the trapping force \cite{Nasyrov(1)01}. Though these more complicated MOTs have been studied, they are still not fully understood.

An exciting recent advance is the laser cooling and magneto-optical trapping of molecules \cite{Shuman(1)10, Barry(1)12, Hummon(1)13, Zhelyazkova(1)14, Barry(1)14}. A 3D MOT of SrF molecules has been demonstrated \cite{Barry(1)14}, as has the compression of a beam of YO molecules using magneto-optical forces \cite{Hummon(1)13}. Several lasers are used to drive multiple transitions, and to avoid branching to other rotational levels one deliberately chooses a cooling transition where the upper state angular momentum is less than that of the lower state. In this case it is challenging, but crucial, to understand the origin of the restoring forces that produce the MOT.

Here, we build a simple, intuitive understanding of how MOTs work in various angular momentum cases, find the correct polarizations for the laser beams in each case, and understand how the signs and magnitudes of the lower and upper-state magnetic g-factors influence the strength of the trapping force and the correct choice of polarization. We use a rate model to calculate the trapping forces for these various cases. We study how a rapid, synchronous reversal of the magnetic field gradient and laser polarizations, as used in \cite{Hummon(1)13}, can increase the trapping forces in some cases. We consider the consequences of our findings for various molecules, model the arrangement used in \cite{Barry(1)14} to make a SrF MOT, and suggest how the confining forces for that MOT can be increased.

\section{Model}\label{Sec:Model}

We consider an atom or molecule of mass $m$ that has a set of lower levels $l$ and upper levels $u$. The angular frequency of the transition from $l$ to $u$ is $\omega_{l,u}$. The upper levels all share a common decay rate $\Gamma$. The atom interacts with several laser beams. Laser beam component $p$ has angular frequency $\omega_{p}$ and propagates in the direction of the unit vector ${\bf k}_p$. The frequencies are all similar, and so we use a single wavelength $\lambda \simeq 2\pi c/\omega_{p}$ for all components. The MOT is made using six orthogonal beams each containing $n_f$ frequencies, and so there are $6n_f$ laser components. The laser beams and magnetic field of the MOT are defined in a laboratory coordinate system whose axes ($x,y,z$) are aligned with the k-vectors of the lasers. The magnetic field is ${\bf B}=A(x {\hat{\bf x}}+y {\hat{\bf y}}-2z {\hat{\bf z}})$ where $A$ is the field gradient in the $xy$-plane. The simulation is done in a local coordinate system whose $z$-axis is always in the direction of the magnetic field. The particles move slowly enough that they adiabatically follow changes in the field direction. The MOT is simulated using rate equations for the position ${\bf r}$ and velocity ${\bf v}$ of the particle, for the populations $N_l$ and $N_u$ of each lower and upper level, and for the number of scattered photons $\gamma$:
\numparts
\begin{eqnarray}
{\bf r'} &= {\bf v}, \label{eq:rprime} \\
{\bf v}'&=\frac{h}{m \lambda}\sum_{l,u,p}{\bf k}_p R_{l,u,p} \left(N_l-N_u\right) + {\bf g}, \label{eq:vprime} \\
N_l'&=\sum_{u,p} R_{l,u,p}\left(N_u-N_l\right) + \sum_u \Gamma r_{l,u}  N_u, \label{eq:Nlprime} \\
N_u'&=-\Gamma  N_u + \sum_{l,p} R_{l,u,p} \left(N_l-N_u\right), \label{eq:Nuprime} \\
\gamma'&=\Gamma \sum_{u} N_{u}. \label{eq:gammaprime}
\end{eqnarray}
\endnumparts
Here, ${\bf g}$ is the acceleration due to gravity, $r_{l,u}$ is the branching ratio for the spontaneous decay of level $u$ to level $l$, and $R_{l,u,p}$ is the excitation rate between levels $l$ and $u$ driven by laser component $p$. This excitation rate is
\begin{equation}
R_{l,u,p}=\frac{\Gamma}{2}\frac{f_{l,u,p} s_{p}}{1+4\left(\delta_{l,u,p}-2\pi{\bf k_{p}\cdot v}/\lambda-\Delta\omega_{l,u}\right)^{2}/\Gamma^{2}},
\end{equation}
where $s_{p}$ is the saturation parameter, $f_{l,u,p}$ is the fractional strength of the transition being driven, $\delta_{l,u,p}=\omega_{p}-\omega_{l,u}$ is the detuning from the resonance angular frequency for a stationary particle at zero field, and $\Delta\omega_{l,u}$ is the Zeeman shift of the transition angular frequency. For the small magnetic fields of a typical MOT we take the Zeeman shift to be linear, $\Delta\omega_{l,u} = (g_u M_u - g_l M_l)\mu_{B}B/\hbar$, where $g_u$, $g_{l}$ are the g-factors and $M_u$, $M_l$ the magnetic quantum numbers of the upper and lower levels. The saturation parameter is $s_{p}=I_p/I_s$, where $I_{p}$ is the intensity of laser component $p$, and $I_{s}=\pi h c \Gamma/(3 \lambda^{3})$ is the saturation intensity for a two-level atom. The transition strength is given by
\begin{equation}
f_{l,u,p} = \frac{\left|\langle l |{\bf d \cdot \epsilon}_{p}|u \rangle\right|^{2}}{\sum_{k} \left|\langle k |{\bf d }|u \rangle\right|^{2}},
\end{equation}
where ${\bf d}$ is the dipole moment operator, ${\bf \epsilon}_p$ is the laser polarization in the local coordinate system, and the sum runs over all lower levels. From these definitions, it follows that $f_{l,u,p}s_{p}=2\Omega_{l,u}^{2}/\Gamma^{2}$, where $\Omega_{l,u}$ is the Rabi frequency at which the transition between $l$ and $u$ is being driven. The polarization of each laser component is specified relative to its own k-vector and is called $\epsilon_{\rm{laser}}$. This is transformed into the local coordinate system via $\epsilon = R(\alpha_{2},\beta_{2},\gamma_{2}).R(-\gamma_1,-\beta_1,-\alpha_1).\epsilon_{\rm{laser}}$, where $R$ is the cartesian rotation matrix, $(\alpha_{1},\beta_{1},\gamma_{1})$ are the Euler angles that rotate the laboratory z-axis onto the k-vector and $(\alpha_{2},\beta_{2},\gamma_{2})$ are the Euler angles that rotate the laboratory z-axis onto the magnetic field vector. The intensity distribution of each laser beam is a truncated Gaussian
\numparts
\begin{eqnarray}
I&=\frac{2P}{\pi w^{2}} \exp(-2 r^{2}/w^{2})\,\,\,\,(r\le r_t),\\
I&=0\,\,\,\,(r>r_t),
\end{eqnarray}
\endnumparts
where $r$ is the distance from the centre of the beam, $w$ is the $1/e^{2}$ radius, $r_{t}$ is a truncation radius, and $P$ is the power of the beam prior to truncation.

For many purposes, such as finding the capture velocity, the trapping frequency, the scattering rate, or the population distribution, the rate equations written above are adequate. However, they cannot give the temperature limit of the MOT since they do not reflect the randomness of the scattering. To capture this, the extra terms $h\Gamma/(m\lambda)({\bf f}_{1}(\gamma)+{\bf f}_{2}(\gamma))\sum_{u}N_{u}$ are added to the right hand side of equation (\ref{eq:vprime}), where ${\bf f}_{1,2}$ are unit vectors whose orientations change randomly each time $\gamma$ increases by 1. These simulate the randomness of the momentum kicks due to absorption and spontaneous emission. With these terms included the integration has to be done in much smaller time steps, and so they are only included when their effects are of interest, e.g. for finding the temperature of the MOT.

It is often useful to compare the simulation results with the results of standard 1D MOT theory \cite{MetcalfBook}. In this theory, the atom has an $F=0$ lower level, and an $F=1$ upper level with magnetic moment $\mu=g_u \mu_B$, and it interacts with two counter-propagating beams of wavevector $k=2\pi/\lambda$ and detuning $\delta_{0}$. The acceleration is
\numparts
\begin{eqnarray}\label{eq:accel1D}
a_{z} &= \frac{\hbar k}{m} (R_{-}-R_{+}),\\
R_{\pm} &= \frac{\Gamma}{2}\frac{s}{1+s+4(\delta_{0} \pm k v_{z} \pm 2\mu A z/\hbar)^{2}/\Gamma^{2}}.
\end{eqnarray}
\endnumparts
For small velocities and displacements, this can be written as $a_{z} = -\beta v_{z} -\omega_{z}^{2} z$, where $\omega_{z}$ is the trapping frequency in the z-direction, given by
\begin{equation}\label{eq:trapFreq1D}
\omega_{z}^{2}=-\frac{16 k s A \mu \delta_{0}}{\Gamma (1+s+4\delta_{0}^{2}/\Gamma^{2})^{2}},
\end{equation}
and $\beta$ is the damping coefficent given by
\begin{equation}
\beta = -\frac{8 s \hbar k^{2} \delta_{0}}{m \Gamma \left(1+s+4\delta_{0}^{2}/\Gamma^{2}\right)^{2}}.
\end{equation}

\section{Angular momentum cases}\label{Sec:AngMomCases}

To build our understanding, we focus on simple cases with just two levels, a lower level of angular momentum $F_{l}$ and an upper level of angular momentum $F_{u}$. We consider an atom at rest and displaced from the centre of the MOT along any one of the principal axes. We refer to the two beams that propagate along this axis as the restoring and anti-restoring beams, where the former is the one whose radiation pressure pushes the atom back towards the centre. The other beams are called the orthogonal beams. All beams have the same frequency and are red-detuned from the atomic transition. In our local coordinate system defined by the magnetic field at the position of the atom, the Zeeman sub-levels of the lower and upper states are labelled $M_{l}$ and $M_{u}$ respectively, and transitions are labelled $\sigma^{\pm}$ according to whether $\Delta M = M_{u}-M_{l}=\pm 1$.

\subsection{$F_{u}=F_{l}+1$}

\begin{figure}[tb]
	\centering
		\includegraphics[width=0.8\textwidth]{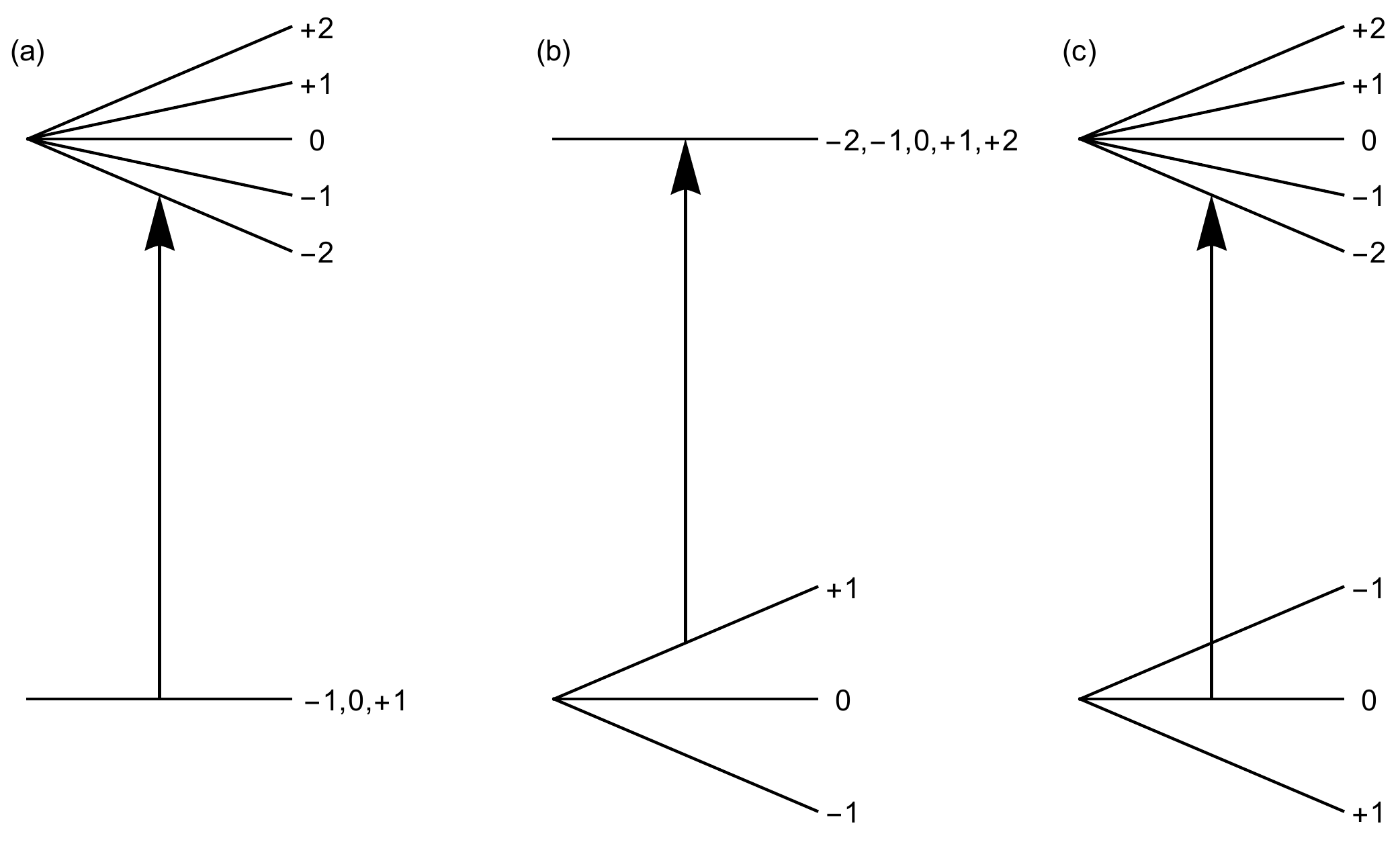}
		\caption{Energy levels as a function of increasing magnetic field for an $F=1$ ground state and an $F=2$ excited state. In (a) the g-factor is zero in the ground state, in (b) it is zero in the excited state, while in (c) it is positive in the excited state and negative in the ground state. The arrow indicates the frequency of red-detuned light.}
		\label{fig:F1toF2}
\end{figure}

First consider the case where $F_{l}=1$ and $F_{u}=2$. This is the simplest case where there cannot be a dark state and where both states can have Zeeman shifts. The restoring beam is polarized to drive $\sigma^{-}$ transitions, while the anti-restoring beam drives $\sigma^{+}$ transitions. Suppose the lower state has no Zeeman splitting, as illustrated in figure \ref{fig:F1toF2}(a). The strongest transition is the one to $M_{u}=-2$ because this is closest to resonance. Only the restoring beam can drive a transition to this state. Atoms are optically pumped into the $M_{l}=-1$ state and then only interact with the restoring beam. This is the way a normal atomic MOT works. The situation is not much changed when both upper and lower states have a positive g-factor. For $M_{l}=0$ nothing is changed, for $M_{l}=+1$ the preference for scattering from the restoring beam is weakened, while for $M_{l}=-1$ the preference is strengthened.

Now consider the case illustrated in figure \ref{fig:F1toF2}(b) where the upper state has no Zeeman splitting. The transition from $M_{l}=+1$ is strongest because it is closest to resonance. Both the restoring and anti-restoring beams can drive transitions out of this state, because the upper state has both $M_{u}=+2$ and $M_{u}=0$ available. The rate for the $\Delta M = +1$ transition is 6 times higher, and so the anti-restoring beam dominates while the atom remains in $M_{l}=+1$. However, following excitation to $M_{u}=0$ the atom can reach $M_{l}=-1$ and then the relative rates are exactly reversed. The fact that the excitation rate out of $M_{l}=-1$ is slower than that out of  $M_{l}=+1$ is not relevant here. The important point is the relative number of photons absorbed from each beam, and this is, on average, equal for the two beams. In this case the ground state with the highest population is $M_{l}=-1$, the excited states all have equal population, and there is no confining force. The argument is unchanged if the polarizations of the lasers are reversed, and the same argument also applies to all other angular momentum cases. If there is no Zeeman splitting in the upper state there is, on average, no selectivity for one laser beam over the other, so there can be no MOT.

Finally consider the case where $g_u$ is positive and $g_l$ is negative. We may wonder whether the polarizations should be reversed in this case. They should not. Let us keep the polarizations the same as before and analyze what happens. For $M_{l}=0$ nothing is changed relative to case (a) and the restoring beam will dominate. For $M_{l}=+1$ the red-detuning once again ensures that the restoring beam dominates, with excitation to $M_{u}=0$ being favoured. For $M_{l}=-1$ the restoring beam will be favoured close to the centre where the field is small and the anti-restoring beam will be favoured further out where the Zeeman shift is large enough to bring the $\sigma^{+}$ transition into resonance. In this last case, the upper state is $M_{u}=0$ which can decay to any of the lower states, and so an atom in $M_{l}=-1$ soon finds itself in one of the other ground states where the restoring beam dominates always.

\begin{figure*}[tb]
	\centering
		\includegraphics[width=\textwidth]{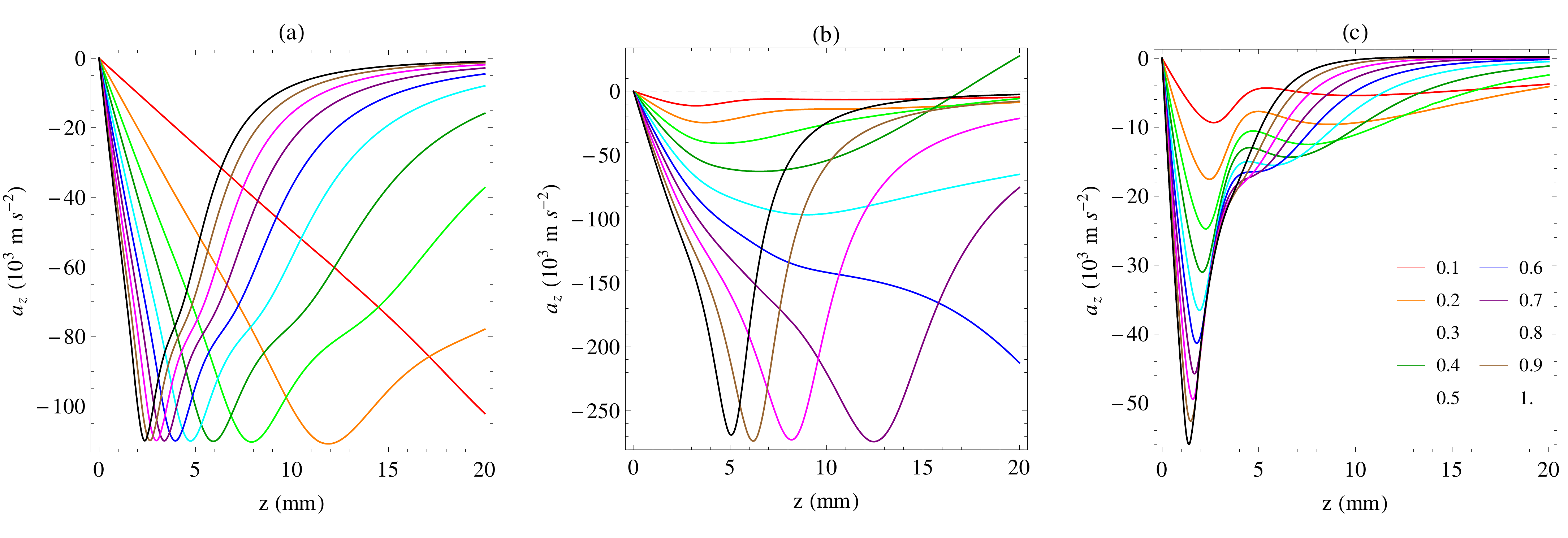}
		\caption{Acceleration versus axial position, for the case where $F_l=1$ and $F_u=2$, and for various values of $g_u$ (colours) and $g_l$: (a) $g_l=0$, (b) $g_l=1$, (c) $g_l=-1$. Parameters are $\lambda = 589$\,nm, $\Gamma=2\pi\times 9.8$\,MHz, $m=23$\,amu, $\delta_0=-\Gamma$, $A=10$\,G/cm, $s=1$.}
		\label{fig:accelCurvesF1toF2}
\end{figure*}

Figure \ref{fig:accelCurvesF1toF2} shows how the acceleration of a stationary atom depends on the axial distance from the MOT centre, for various values of $g_{l}$ and $g_{u}$. These results are calculated using the model described in section \ref{Sec:Model} applied to this angular momentum case. The mass, linewidth and transition wavelength are those of $^{23}$Na, the radial field gradient is $A=10$\,G/cm, the detuning is $\delta_{0}=-\Gamma$, the intensity is $s=1$, the laser beams are $\sigma^{-}$ polarized, and the beams are made sufficiently large that the intensity is nearly uniform across the whole distance range of interest. Figure \ref{fig:accelCurvesF1toF2}(a) shows the acceleration profile in the case where $g_{l}=0$, and for various values of $g_{u}$. As the atom moves away from the centre $|a_{z}|$ first increases linearly, then reaches a maximum at the point where the Zeeman shift of the $M_{u}=-2$ level is approximately equal to the detuning, then gradually decreases. As $g_{u}$ is increased, the turning point moves inwards in proportion, but the maximum value of $|a_{z}|$ does not change. One finds the same dependence on the field gradient $A$, because it is the product $g_{u} A$ that matters. As expected, these profiles are similar to those obtained from standard 1D MOT theory, equation (\ref{eq:accel1D}). The main difference is that the maximum value of $|a_{z}|$ in the figure is 3.6 times smaller than found from this equation. That is because the orthogonal laser beams partly saturate the transition, leading to a reduction in the scattering rate from the restoring beam, and because the orthogonal beams can drive $\Delta M = 0$ transitions that pump population out of the $M_{l}=-1$ state which has the highest scattering rate.

Figure \ref{fig:accelCurvesF1toF2}(b) shows the acceleration profile in the case where $g_{l}=+1$ and for various values of $g_{u}$. As argued above, there is no trapping when the upper level has no Zeeman shift, and similarly the acceleration is small when $g_{u}$ is small. As $g_{u}$ increases towards 0.6 there is an increasing preference to drive $\Delta M=-1$ transitions, and so $|a_{z}|$ increases and its maximum value shifts to larger $z$. For $g_{u}>0.6$ the maximum acceleration is nearly constant, and the position of the maximum shifts inwards, similar to the behaviour seen for $g_{l}=0$. Note that the maximum acceleration is about 2.5 times larger than in case (a). At the value of $B$ that brings the $-1 \rightarrow -2$ transition into resonance, the Zeeman splitting of the upper levels is $\Delta E_{\rm{res}}=-\delta_0 g_u/(2g_u - g_l)$. When $g_{l}=0$ this is always $-\delta_{0}/2$, but when $g_l=1$ and $g_u=0.6$ it is $-3\delta_{0}$. This much larger splitting produces a stronger preference for scattering from the restoring beam, and therefore a stronger force.

Figure \ref{fig:accelCurvesF1toF2}(c) shows the acceleration profile in the case where $g_{l}=-1$ and for various values of $g_{u}$. In this case the maximum $|a_{z}|$ increases with $g_{u}$ over the whole range explored, because for $g_{l}=-1$, $\Delta E_{\rm{res}}$ increases with $g_u$. It tends towards $\Delta E_{\rm{res}}=-\delta_{0}/2$ in the limit of large $g_u$, and so the maximum acceleration tends towards the value in case (a). Note that increasing $g_u$ also narrows the range of $z$ over which the acceleration is significant, reducing the MOT capture radius.

Near the centre of the MOT the acceleration depends linearly on $z$. Its gradient is the square of the trapping frequency, $\omega_{z}^{2}$, and this is found to be independent of $g_l$. In 1D MOT theory, the trapping frequency is given by equation (\ref{eq:trapFreq1D}). The actual value of $\omega_z$ is reduced because of the effects of the orthogonal beams, but the numerical model shows that its dependence on $\delta_{0}$, $s$, and $g_u A$ follows closely this standard expression. This is also true for the other angular momentum cases considered.

\subsection{$F_{u}=F_{l}$}\label{sec:F1F1}

\begin{figure}[tb]
	\centering
		\includegraphics[width=0.6\textwidth]{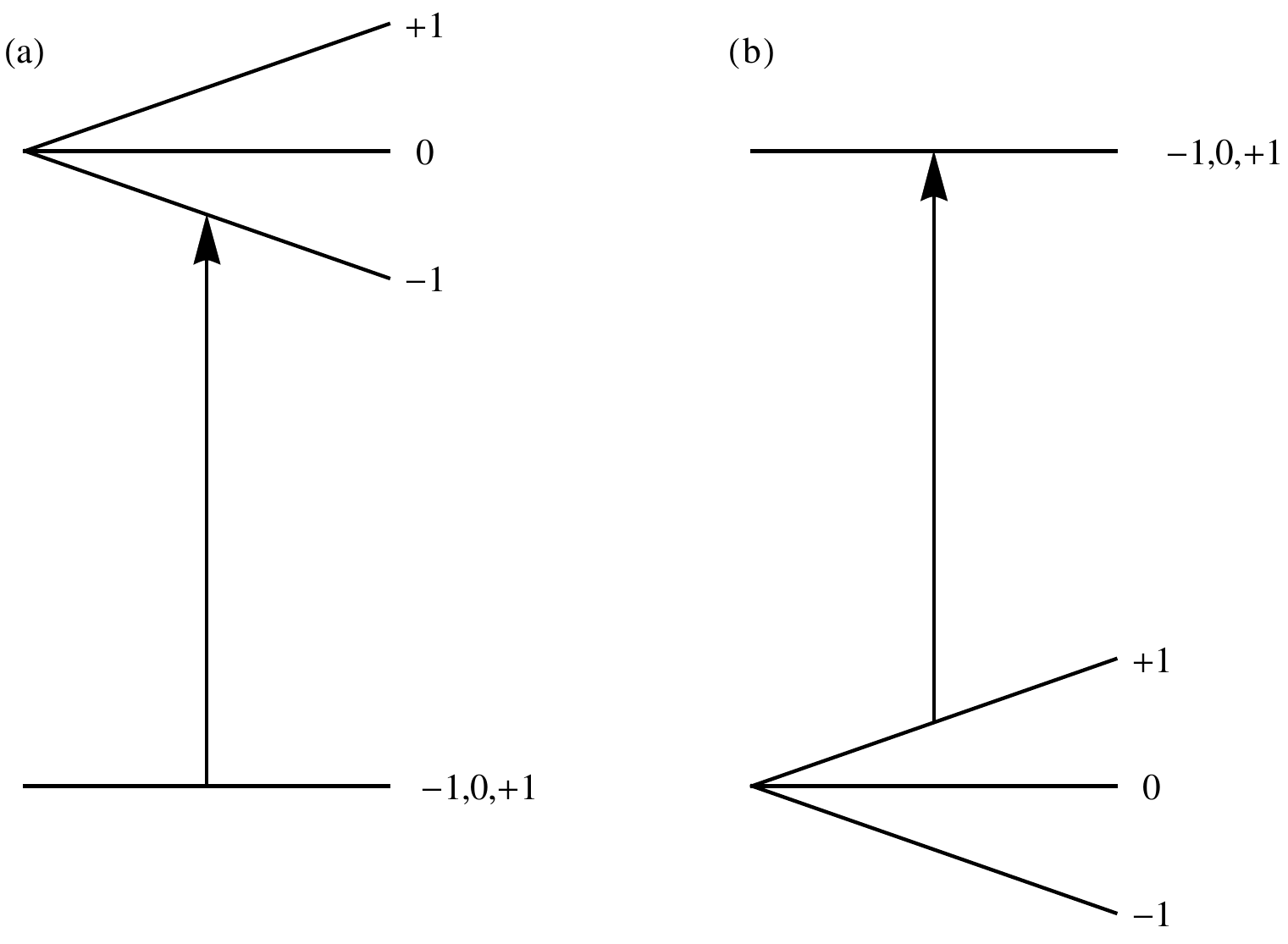}
		\caption{Energy levels as a function of increasing magnetic field for an $F=1$ ground state and an $F=1$ excited state. In (a) the g-factor is zero in the ground state, while in (b) it is zero in the excited state. The arrow indicates the frequency of red-detuned light.}
		\label{fig:F1toF1}
\end{figure}

Next, consider the case where both ground and excited states have angular momentum $F=1$, and where the lower state has no Zeeman shift, as illustrated in figure \ref{fig:F1toF1}(a). The polarizations of the laser beams are the same as before. Suppose that there are no orthogonal laser beams. Then, the transition out of $M_{l}=+1$ can only be driven by the restoring beam, while the transition out of $M_{l}=-1$ can only be driven by the anti-restoring beam. In both cases the excited state is $M_{u}=0$ which decays with equal probability to each of $M_{l}=\pm 1$. It follows that the mean number of photons scattered from each beam is equal and there is no net force. The $M_{l}=0$ state plays no part, apart from an initial transient where population is pumped out of this state. The introduction of the orthogonal beams changes this picture. An atom in $M_{l}=+1$ can interact with either the restoring beam or the orthogonal beams, with a preference for exciting to $M_{u}=0$ which is closer to resonance. The restoring beam drives this transition more strongly than any one of the orthogonal beams, but there are four orthogonal beams so it tends to even out. An atom in $M_{l} = 0$ can interact with any of the beams, but is more likely to interact with the restoring beam than the anti-restoring beam because the $\Delta M =-1$ transition is closest to resonance. An atom in $M_{l}=-1$ can interact with the anti-restoring beam or the orthogonal beams, but only the orthogonal beams can drive the transition close to resonance and so they dominate. To complete the picture we need to know which of the three ground states is visited most often. There is a tendency to excite to $M_{u}=-1$, which then decays equally to $M_{l}=0,-1$. When the Zeeman splitting and detuning are large, nearly half of all scattering events are driven from $M_{l}=-1$ by the orthogonal beams, while the other half are from $M_{l}=0$ and are sometimes driven by the restoring beam but rarely by the anti-restoring beam. There is a net restoring force. For smaller Zeeman splittings this trend applies more weakly, but in addition the $M_{l}=+1$ state is sometimes visited and transitions out of this state can be driven by the restoring but not by the anti-restoring beam.

Now consider what happens when only the lower state has a Zeeman shift, as illustrated in figure \ref{fig:F1toF1}(b). When $M_{l}=+1$ transitions are driven by either the restoring or orthogonal beams. When $M_{l}=0$ transitions can be driven by any of the beams, and the scattering rates for the restoring and anti-restoring beams are equal. When $M_{l}=-1$ transitions are driven by either the anti-restoring or orthogonal beams. Each upper state decays to two lower states with equal probability and so the $M_{l}=\pm 1$ states are visited an equal number of times. It follows that there is no net force.

It is possible to recover a net force however. The population tends to accumulate in $M_{l}=-1$ because the excitation rate out of this state is slow, but the restoring force comes only from transitions out of $M_{l}=+1$. A net restoring force can be produced by rapidly switching the population back into $M_{l}=+1$ each time it starts to build up in $M_{l}=-1$. This is equivalent to a synchronous reversal of the magnetic field gradient and laser beam polarizations, provided they are switched rapidly enough that the atomic state cannot follow the change. This is discussed in more detail in section \ref{Sec:RapidSwitch}.

\begin{figure*}[tb]
	\centering
    \includegraphics[width=\textwidth]{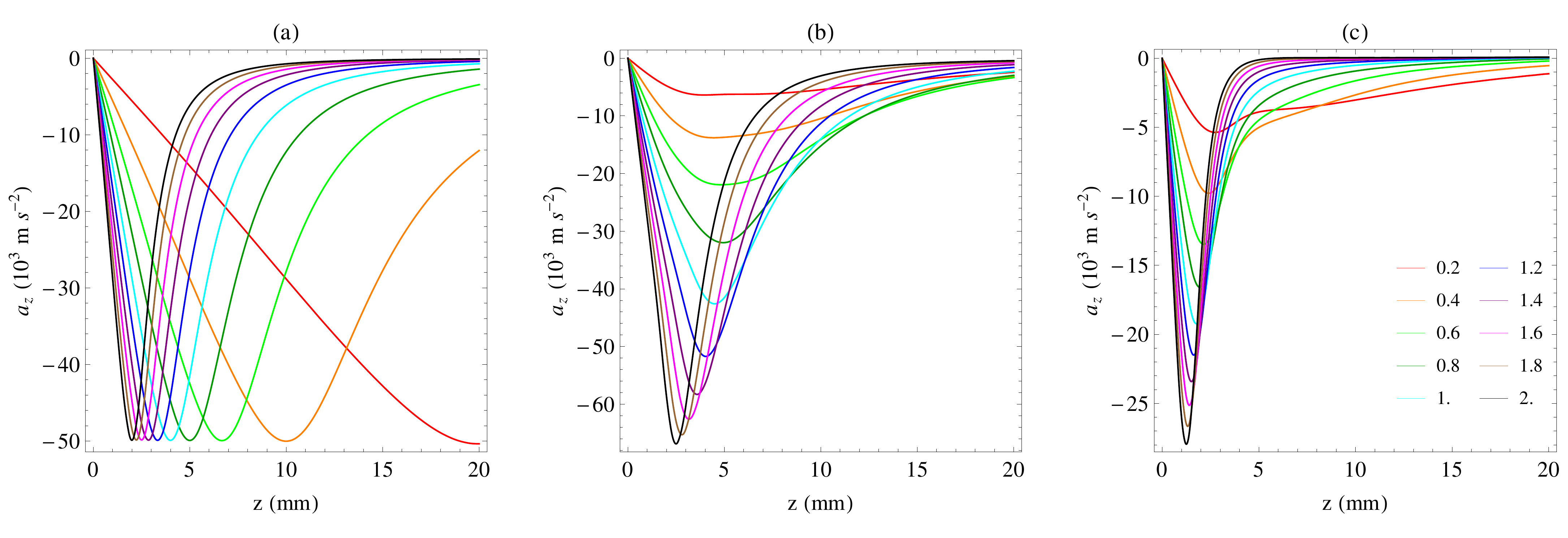}
		\caption{Acceleration versus axial position, for the case where $F_l=1$ and $F_u=1$, and for various values of $g_u$ (colours) and $g_l$: (a) $g_l=0$, (b) $g_l=1$, (c) $g_l=-1$. Parameters are $\lambda = 589$\,nm, $\Gamma=2\pi\times 9.8$\,MHz, $m=23$\,amu, $\delta_0=-\Gamma$, $A=10$\,G/cm, $s=1$.}
		\label{fig:accelCurvesF1toF1}
\end{figure*}

Figure \ref{fig:accelCurvesF1toF1} shows how the acceleration depends on axial position for this angular momentum case, and for various values of $g_u$ and $g_l$. These curves, and the trends they display, are similar to those in figure \ref{fig:accelCurvesF1toF2} which was discussed above. The main differences are featured in case (b), where $g_l=1$. Here, when $g_u<g_l$, the restoring force comes mainly from the $+1\rightarrow 0$ transition at low field, and mainly from the $0 \rightarrow -1$ transition at higher field. As $g_u$ increases, the latter transition comes into resonance for smaller values of $B$ and so the peak of $|a_{z}|$ moves inwards towards the trap centre. The maximum obtainable trapping force is reduced by a factor of about 4 relative to figure \ref{fig:accelCurvesF1toF2}(b), because the atoms cannot continuously scatter photons from the restoring beam without being pumped into a state that is dark to that beam.

We found that in 1D the average force is zero when $F_u=F_l=1$. This is special to that case, and is not true of other cases where $F_u=F_l$. Consider, for example, a 1D MOT with $F_u=F_l=2$. Because the light is red-detuned, the restoring beam is favoured for all $M_{l}$ apart from $M_{l} = -2$. An atom will be optically pumped to $M_{l}=-2$ by the restoring beam, and once in this state the anti-restoring beam will drive the transition to $M_{u}=-1$. That upper state decays to $M_{l}=-2,-1,0$ with probabilities 1/3, 1/6 and 1/2. Because these probabilities are biased away from $M_{l}=-2$, it takes, on average, more scattering events to reach this state than it does to leave it. So there is a restoring force, even in 1D. In 3D, we find that the trapping force for $F_u=F_l=2$ is roughly half that found for $F_u=F_l=1$, and higher values of $F$ give even weaker forces, but otherwise the characteristics are similar.

\subsection{$F_{u}=F_{l}-1$}

\begin{figure}[tb]
	\centering
		\includegraphics[width=0.6\textwidth]{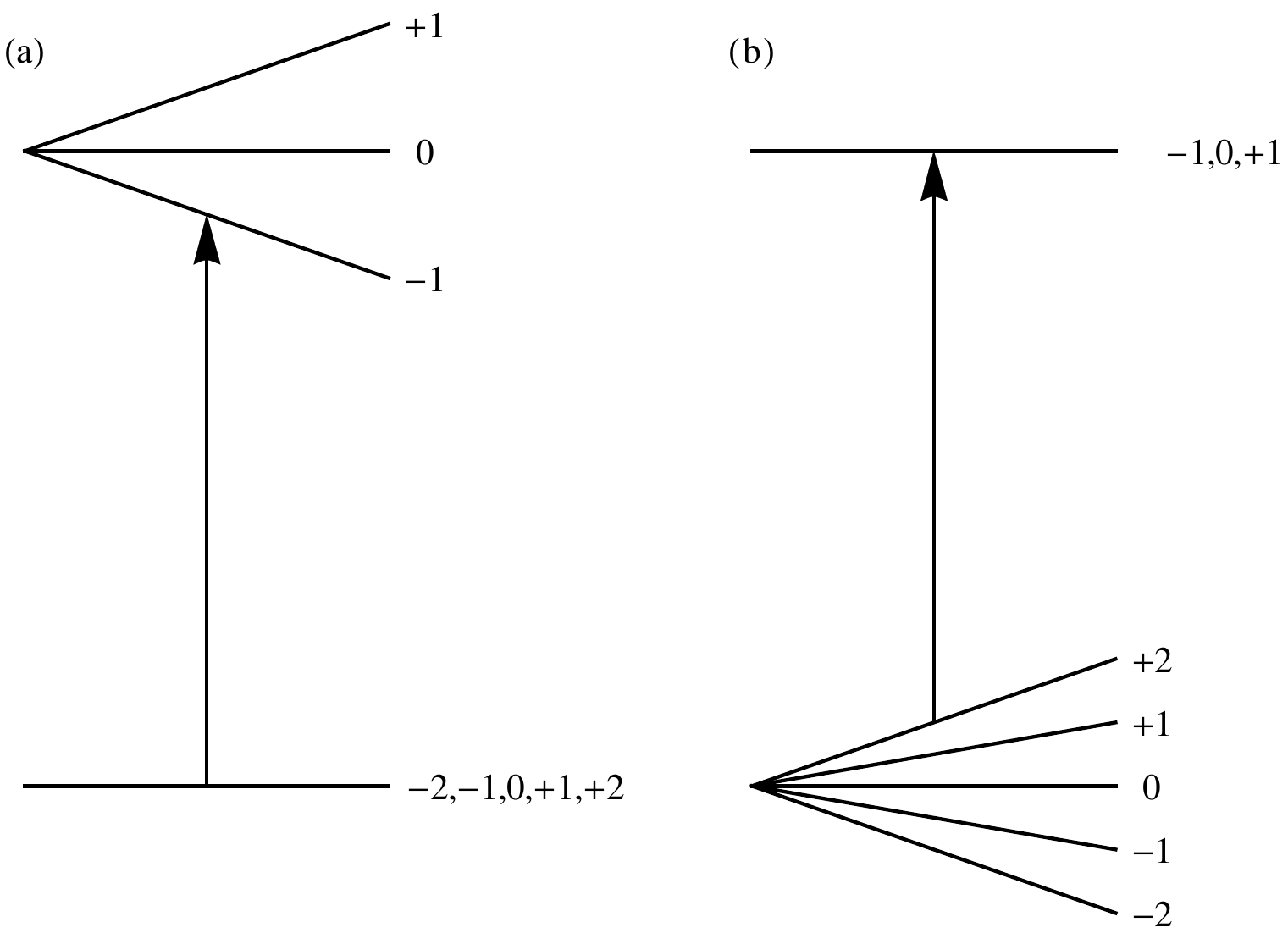}
		\caption{Energy levels as a function of increasing magnetic field for an $F=2$ ground state and an $F=1$ excited state. In (a) the g-factor is zero in the ground state, while in (b) it is zero in the excited state. The arrow indicates the frequency of red-detuned light.}
		\label{fig:F2toF1}
\end{figure}

Consider now the case illustrated in figure \ref{fig:F2toF1} where the ground state has angular momentum $F_{l}=2$ and the excited state has $F_{u}=1$. Once again, suppose there is no Zeeman splitting in the ground state. The transition to $M_{u}=-1$ is driven most strongly because this transition is closest to resonance. This upper state decays 60\% of the time to $M_{l} = -2$, 30\% of the time to $M_{l}=-1$, and 10\% to $M_{l}=0$. The combination of the resonance condition and the branching ratios provide a strong tendency to optically pump to the negative $M_{l}$ states, particularly to $M_{l}=-2$. In fact, when the Zeeman splitting is large, the atom hardly ever reaches the positive $M$ states, and 90\% of all scattering events are from $M_{l}=-2$ or $M_{l}=-1$. In this case, it must be advantageous for the restoring beam to drive $\sigma^{+}$ transitions, requiring the the opposite polarization to the other cases considered. Let's analyze in more detail what happens for each ground state with polarizations chosen this way. When $M_{l}=+2,+1,0$ the anti-restoring and orthogonal lasers tend to drive $\sigma^{-}$ transitions, since these are closer to resonance, while the restoring laser does nothing. When $M_{l}=-1$ the orthogonal lasers tend to drive the transition to $M_{u}=-1$. The restoring and orthogonal lasers can drive transitions to $M_{u}=0$, but more weakly. When $M_{l}=-2$ the orthogonal and restoring lasers drive the only possible transition. We see that for 3 out of the 5 lower states the anti-restoring beam dominates over the restoring beam. In the other 2 states the restoring beam dominates over the anti-restoring beam, although in one of these most scattering is from the orthogonal beams. Nevertheless, the net effect is a restoring force because the $M_{l}=-2,-1$ states are visited more often than the others.

Now look at the case where there is no Zeeman shift in the upper state, as shown in figure \ref{fig:F2toF1}(b), and the laser polarizations are still reversed as above. When $M_{g}$ is positive the restoring beam cannot act, when $M_{g}$ is negative the anti-restoring beam cannot act, and when $M_{g} = 0$ the scattering rates from the two beams are equal. The positive and negative $M_{g}$ are visited equally on average, since there is nothing to break the symmetry between them. So there is no net force. This is true for any choice of beam polarization.

\begin{figure*}[tb]
	\centering
		\includegraphics[width=\textwidth]{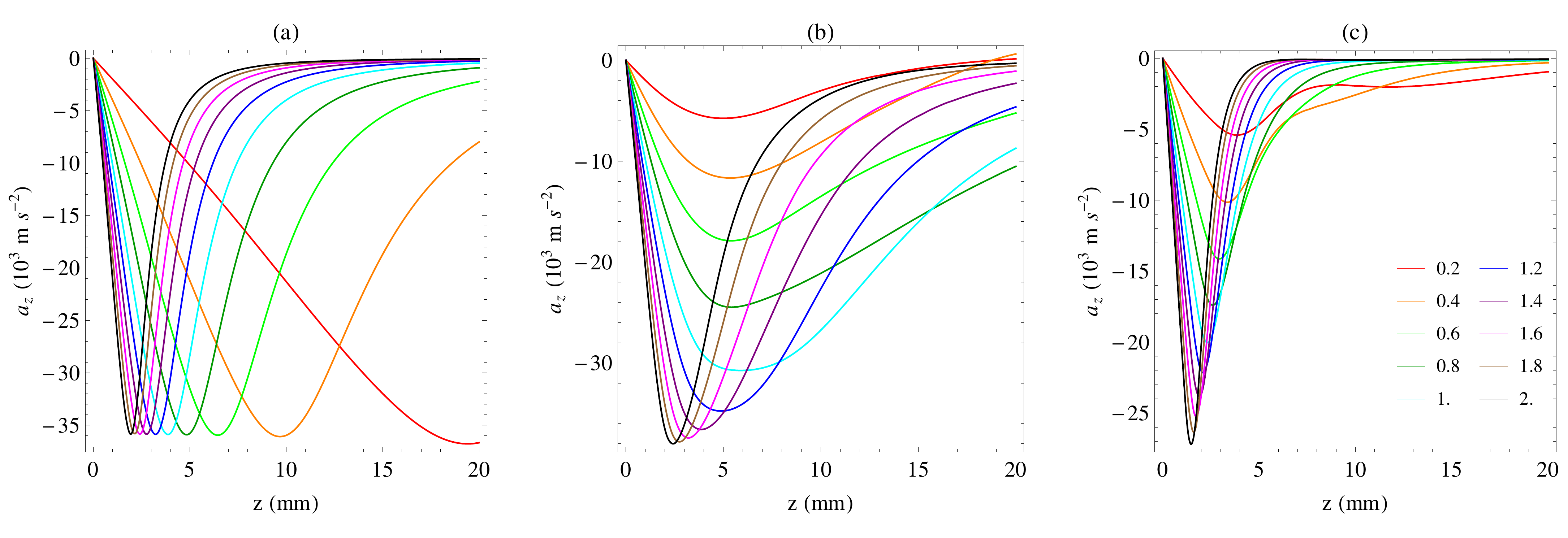}
		\caption{Acceleration versus axial position, for the case where $F_l=2$ and $F_u=1$, and for various values of $g_u$ (colours) and $g_l$: (a) $g_l=0$, (b) $g_l=0.5$, (c) $g_l=-0.5$. Parameters are $\lambda = 589$\,nm, $\Gamma=2\pi\times 9.8$\,MHz, $m=23$\,amu, $\delta_0=-\Gamma$, $A=10$\,G/cm, $s=1$.}
		\label{fig:accelCurvesF2toF1}
\end{figure*}

Figure \ref{fig:accelCurvesF2toF1} shows how the acceleration depends on the axial position in the case where $F_l=2$ and $F_u=1$, with the restoring beams driving $\sigma^{+}$ transitions. These curves are similar to those discussed for the other two angular momentum cases.

\subsection{Summary}

We have seen that the strength of the restoring force in the MOT depends on the angular momenta of the upper and lower states and on the ratio of their Zeeman shifts. The restoring forces are weak when the Zeeman shift of the upper state is small compared to that of the lower state. The correct choice of polarization depends on the sign of the g-factor in the upper state, but not on its sign in the lower state. Table \ref{tab:accelSummary} summarizes the correct choice of polarization for positive $g_{u}$ and for various angular momentum cases, and in each case gives the maximum force found for zero, positive, and negative values of $g_l$. When $g_{u}$ is negative the polarizations should be reversed. It is interesting to compare the polarization rules summarized in table \ref{tab:accelSummary} with experimental results. In \cite{Flemming(1)97}, a sodium MOT operating on the D$_1$ line was reported, using two laser frequencies to drive transitions from both ground state hyperfine levels. MOTs were produced using four distinct configurations for the frequencies and polarizations of the lasers (see figure 2 of \cite{Flemming(1)97}), but no explanation was given as to why these configurations are the correct ones. All four configurations conform to the polarization rules of \ref{tab:accelSummary}.

Note that the magneto-optical forces are unchanged if the detuning and polarizations are both reversed. A blue detuning heats the atoms and so is usually undesirable, but when there are closely spaced hyperfine components a blue detuning relative to some of the transitions may be inevitable, and this observation is then relevant.

\begin{table}[tb]
\begin{center}
\begin{tabular}{|c|c|c|c|c|c|}
\hline
$F_{l}$ & $F_{u}$ & Pol. & \multicolumn{3}{|c|}{$a_{\rm{max}}/a_{\rm{max}}^{\rm{1D}}$ (\%)} \\
\hline
& & & (i) & (ii) & (iii) \\
\hline
0 & 1 & $\sigma^{-}$ & 70 & -- & -- \\
1 & 2 & $\sigma^{-}$ & 28 & 68 & 20\\
2 & 3 & $\sigma^{-}$ & 21 & 55 & 12 \\
1 & 1 & $\sigma^{-}$ & 13 & 17 & 7.0 \\
2 & 2 & $\sigma^{-}$ & 6.0 & 10 & 3.8 \\
3 & 3 & $\sigma^{-}$ & 4.2 & 8.0 & 2.7 \\
2 & 1 & $\sigma^{+}$ & 9.0 & 8.5 & 6.8 \\
3 & 2 & $\sigma^{+}$ & 11 & 16 & 7.4 \\
\hline
\end{tabular}
\end{center}
\caption{
\label{tab:accelSummary}
Required polarization and maximum acceleration in a MOT with lower and upper angular momenta $F_l$ and $F_u$. The notation $\sigma^{\pm}$ means that, in a coordinate system with $z$-axis in the magnetic field direction, the restoring beams drive $\Delta M=\pm 1$ transitions. The polarizations are the correct ones for positive $g_{u}$. When $g_u$ is negative, they should be reversed. Parameters used in the numerical simulation are are $\lambda = 589$\,nm, $\Gamma=2\pi\times 9.8$\,MHz, $m=23$\,amu, $\delta_0=-\Gamma$, $A=10$\,G/cm, $s=1$. The acceleration is divided by the maximum acceleration obtained from equation (\ref{eq:accel1D}), and is given for three cases: (i) $g_{l}=0$, (ii) $g_{l}F_{l}=1, g_{u}F_{u}=2$  (iii)  $g_{l}F_{l}=-1, g_{u}F_{u}=2$.
}
\end{table}

\section{Rapid switching}\label{Sec:RapidSwitch}

\begin{figure*}[tb]
	\centering
		\includegraphics[width=\textwidth]{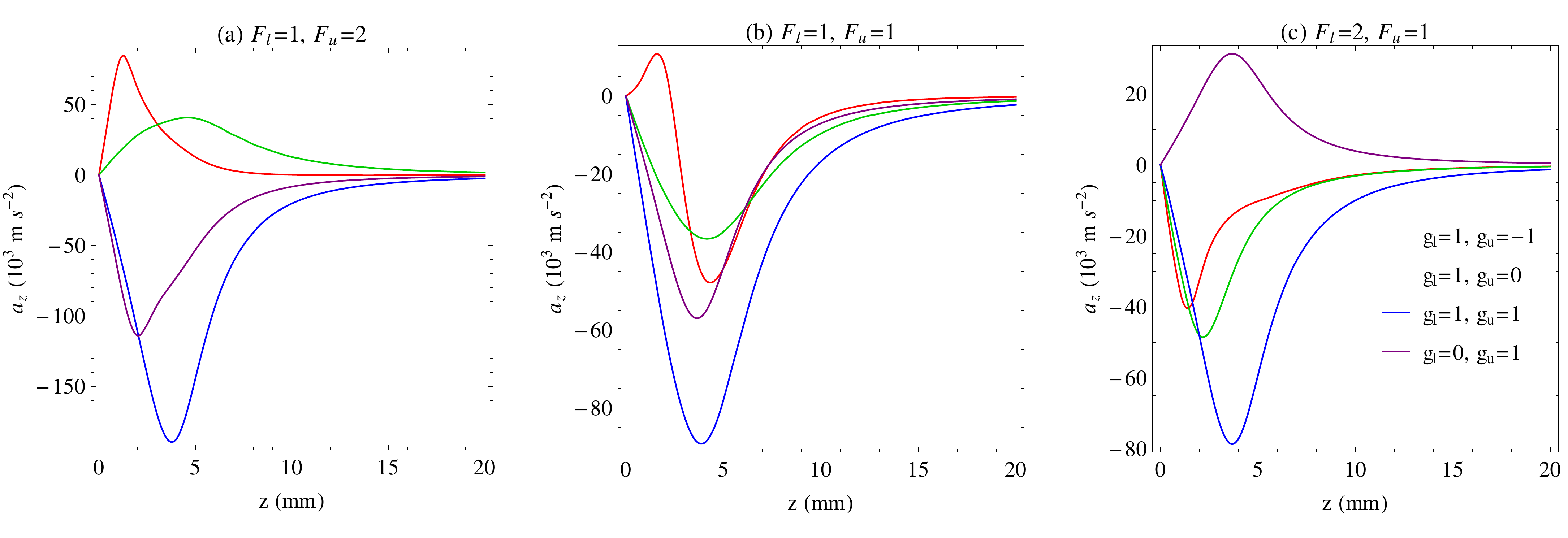}
		\caption{Acceleration versus axial position for various angular momentum cases, when rapid switching is used. The colours correspond to various choices of $g_{l}$ and $g_{u}$, as indicated in the legend. Reversing the signs of both $g_{u}$ and $g_{l}$ reverses the sign of the force. Parameters are $\lambda = 589$\,nm, $\Gamma=2\pi\times 9.8$\,MHz, $m=23$\,amu, $\delta_0=-\Gamma$, $A=10$\,G/cm, $s=1$, $\tau=0.25\,\mu$s. In all cases the polarizations are chosen so that the restoring beam drives $\Delta M=-1$ transitions.}
		\label{fig:accelCurvesRapidSwitching}
\end{figure*}

In those cases where the confining force is weak, e.g. due to a small upper state Zeeman splitting, a stronger confining force may be produced by rapidly switching the magnetic field gradient and laser beam polarizations, as done in reference \cite{Hummon(1)13}. The timescale of the switch is too fast for the atomic state to follow adiabatically. In our rate model, and our local coordinate system where the $z$-axis is always in the magnetic field direction, this switch is equivalent to swapping the populations of sub-levels $M$ and $-M$ of every level. The simulation handles the switch by swapping these populations at each time $t=n \tau$, where $n$ is an integer and $\tau$ is half the switching period.

This method of rapid switching was discussed briefly in section \ref{sec:F1F1}, for the case where $F_l=F_u=1, g_{u}=0$. This is the easiest case to understand. Atoms are optically pumped into $M_{l}=-1$, because the transition out of this state is furthest from resonance, but the switch transfers it to $M_{l}=+1$ and the transition out of this state is driven promptly by the restoring beam. The acceleration curve for this case is shown by the green line in figure \ref{fig:accelCurvesRapidSwitching}(b), where we have chosen $\tau=0.25\,\mu$s. A moderately strong trapping force is obtained. This force gets stronger for shorter values of $\tau$, until $\tau$ is shorter than the optical pumping time. The other curves in figure \ref{fig:accelCurvesRapidSwitching}(b) show what happens for various other values of $g_{u}$ and $g_{l}$. In every case shown, there is a tendency to optically pump towards $M_{l}=-1$, but the trapping force is strongest when $M_{l}=+1$, and so the rapid switching can increase the confining force. For example, when $g_{l}=g_{u}=1$ and $\tau=0.25\,\mu$s, the maximum value of $|a_z|$ is roughly doubled relative to the dc case (compare figures \ref{fig:accelCurvesF1toF1}(b) and \ref{fig:accelCurvesRapidSwitching}(b)).

Figure \ref{fig:accelCurvesRapidSwitching}(a) shows the acceleration curves for the case $F_{l}=1, F_{u}=2$, with the restoring beam driving $\sigma^{-}$ transitions. Here, when $g_{u}=0$, the rapid switching produces an anti-confining force, as shown by the green curve. The opposite choice of polarization gives the confining force. As discussed in the context of figure \ref{fig:F1toF2}(b), the atoms are optically pumped towards $M_{l}=-1$. The switch brings them to $M_{l}=+1$, and from this level the $\Delta M=+1$ transition is 6 times stronger than the $\Delta M=-1$ transition. So, with rapid switching, the restoring beam should be polarized to drive $\Delta M=+1$ transitions. In the cases where $g_u$ and $g_l$ have similar magnitudes the dc MOT already works well and the rapid switching does not help.

Figure \ref{fig:accelCurvesRapidSwitching}(c) shows the acceleration curves for the case $F_{l}=2, F_{u}=1$, with the restoring beam driving $\sigma^{-}$ transitions. When $g_{u}=0$ and $g_{l}$ is positive (as in figure \ref{fig:F2toF1}(b)), atoms are optically pumped towards the negative $M_{l}$ states. The switch transfers them to the positive $M_{l}$ states, and transitions out of these states can be driven by the restoring beam, but not by the anti-restoring beam, so the atoms are confined. In the case where $g_{l}=0$ and $g_{u}$ is positive, the opposite polarization is needed for the same reason as in the dc case.

The results for other combinations of $g_l$ and $g_u$ can be found from the curves in figure \ref{fig:accelCurvesRapidSwitching} by noting that the sign of the force is reversed if the signs of both $g_u$ and $g_l$ are reversed.

\section{Application to molecules with a $^{2}\Sigma - ^{2}\Pi_{1/2}$ cooling transition}

Many molecules are potentially amenable to laser cooling and trapping in a MOT. For those molecules that have so far been cooled \cite{Shuman(1)10, Barry(1)12, Hummon(1)13, Zhelyazkova(1)14, Barry(1)14}, the cooling transition is from X$\,^{2}\Sigma^{+} (N=1)$ to A$^{2}\Pi_{1/2}(J=1/2,p=+1)$, where $N$ is the rotational quantum number, $J$ is the total electronic angular momentum, and $p$ is the parity. The g-factor of a $^{2}\Pi_{1/2}$ state is typically small, which is problematic for making a MOT, as discussed above. In a magnetic field $B_{Z}$, the Zeeman interaction energy for a Hund's case (a) state, $|\Lambda;S,\Sigma;J,\Omega,M\rangle$, is
\begin{equation}
\Delta E_{z} = \mu_{B} B_{Z} \frac{g_{L}\Lambda + g_{S}\Sigma}{J(J+1)}\Omega M,
\end{equation}
where $\mu_{B}$ is the Bohr magneton, $g_{L}=1$, and $g_{S}=2.0023$. For a $^{2}\Pi_{1/2} (J=1/2)$ state, which has $\Lambda= 1$, $\Sigma=-1/2$, $J=1/2$, $\Omega=1/2$, we find $\Delta E_{z} = 7.7 \times 10^{-4} \mu_{B} B_{Z} M$. We see that the g-factor is very close to zero because the magnetic moments associated with the spin and orbital angular momenta are almost equal in magnitude and opposite in direction. The contribution from the magnetic moments associated with the nuclear spin and the rotation of the molecule are also of order $10^{-3}$. The lower level of the cooling transition has $g_{l} \sim 1$. Such a small ratio of $g_u$ to $g_l$ would lead to extremely weak trapping forces in the MOT, unless the method of rapid polarization modulation is used.

In reality, the excited state is not usually a pure $^{2}\Pi_{1/2} (J=1/2)$ state, but is typically mixed by rotational and spin-orbit interactions with a nearby $^{2}\Sigma_{1/2}$ state (the B state for all the molecules listed in Table \ref{tab:gfacs}). The $^{2}\Pi-^{2}\Sigma$ mixing introduces additional terms into the effective Zeeman Hamiltonian \cite{BC}. For the molecules in Table \ref{tab:gfacs}, the most important additions to the effective Hamiltonian are the two parity-dependent terms, which have coefficients known as $g_{l}'$ and $g_{r}^{e'}$. Their matrix elements in a case (a) basis are given in equation (9.71) of reference \cite{BC}. Evaluating these, we find that for a $^{2}\Pi_{1/2}(J=1/2,p=\pm 1,F=1)$ state, the parity-dependent part of the Zeeman shift is $\Delta E_{Z\pm} = \pm \frac{1}{3}(g_{l}'-g_{r}^{e'})\mu_{B} B_{Z} M_{F}$, where the sign refers to the parity of the $\Lambda$-doublet component. The two coefficients are not often measured. However, the mixing that leads to the parity-dependent Zeeman shift also leads to the $\Lambda$-doubling itself, and so there are relationships between the Zeeman coefficients ($g_{l}'$ and $g_{r}^{e'}$) and the $\Lambda$-doubling coefficients ($p$ and $q$). These relationships are $g_{l}'=p/2B$ and $g_{r}^{e'}=-q/B$ \cite{Curl(1)65, Brown(1)78}, where $B$ is the rotational constant.

\begin{table}[tb]
\begin{center}
\begin{tabular}{c|c|c|c|c}
\hline
Molecule & $B$ (cm$^{-1}$) & $p+2q$ (cm$^{-1}$) & $\frac{1}{3}(g_{l}'-g_{r}^{e'})$ & Reference\\
\hline
CaF &  0.34748 & -0.0439 & -0.0211 & \cite{Nakagawa(1)78}\\
SrF & 0.25135 & -0.13291 & -0.088 & \cite{Steimle(1)78} \\
YO & 0.385785 & -0.15061 & -0.065 & \cite{Bernard(1)83} \\
YbF & 0.247966 & -0.3979 & -0.268 & \cite{Steimle(1)07} \\
\hline
\end{tabular}
\end{center}
\caption{
\label{tab:gfacs}
The parity-dependent g-factor of the A$^{2}\Pi_{1/2}(v=0,J=1/2)$ state of a few relevant molecules, inferred from the measured $\Lambda$-doubling coefficient $p+2q$ and rotational constant $B$.
}
\end{table}

Table \ref{tab:gfacs} gives the value of $\Delta E_{Z\pm}/(\mu_{B} B_{Z} M_{F})$ inferred from the measured values of $p+2q$ and $B$, for a few relevant molecules. For CaF, the Zeeman-splitting in the A$^{2}\Pi_{1/2}(J=1/2,F=1)$ excited state is about 40 times smaller than in the X\,$^{2}\Sigma^{+}(N=1,F=2)$ ground state, which is far too small for a dc MOT to be feasible using this transition. Rapid polarization switching can be used to avoid this problem, as discussed in section \ref{Sec:RapidSwitch}. Alternatively, the B\,$^{2}\Sigma^{+}(v=0)$ state can be used for the main cooling transition, so that the upper and lower g-factors are comparable. The Franck-Condon factor between B$(v=0)$ and X$(v=0)$ is very close to 1, and the transition from B to A, which would take molecules out of the cooling cycle, is strongly suppressed because both the transition dipole moment and the $\omega^{3}$ factor are small. This cooling scheme also has the advantage that each vibrational level of X can be driven to a different upper state, which increases the overall scattering rate (see, for example, the appendix of \cite{Tar13a}).

For SrF and YO the excited state Zeeman splitting is about a tenth of the ground state splitting, good enough to make a weakly confining MOT without using the rapid switching method. The trapping forces should be stronger again for YbF, whose excited state Zeeman splitting is about a quarter of the ground state splitting.

\section{Simulating a SrF MOT}

\begin{figure*}[tb]
	\centering
		\includegraphics[width=\textwidth]{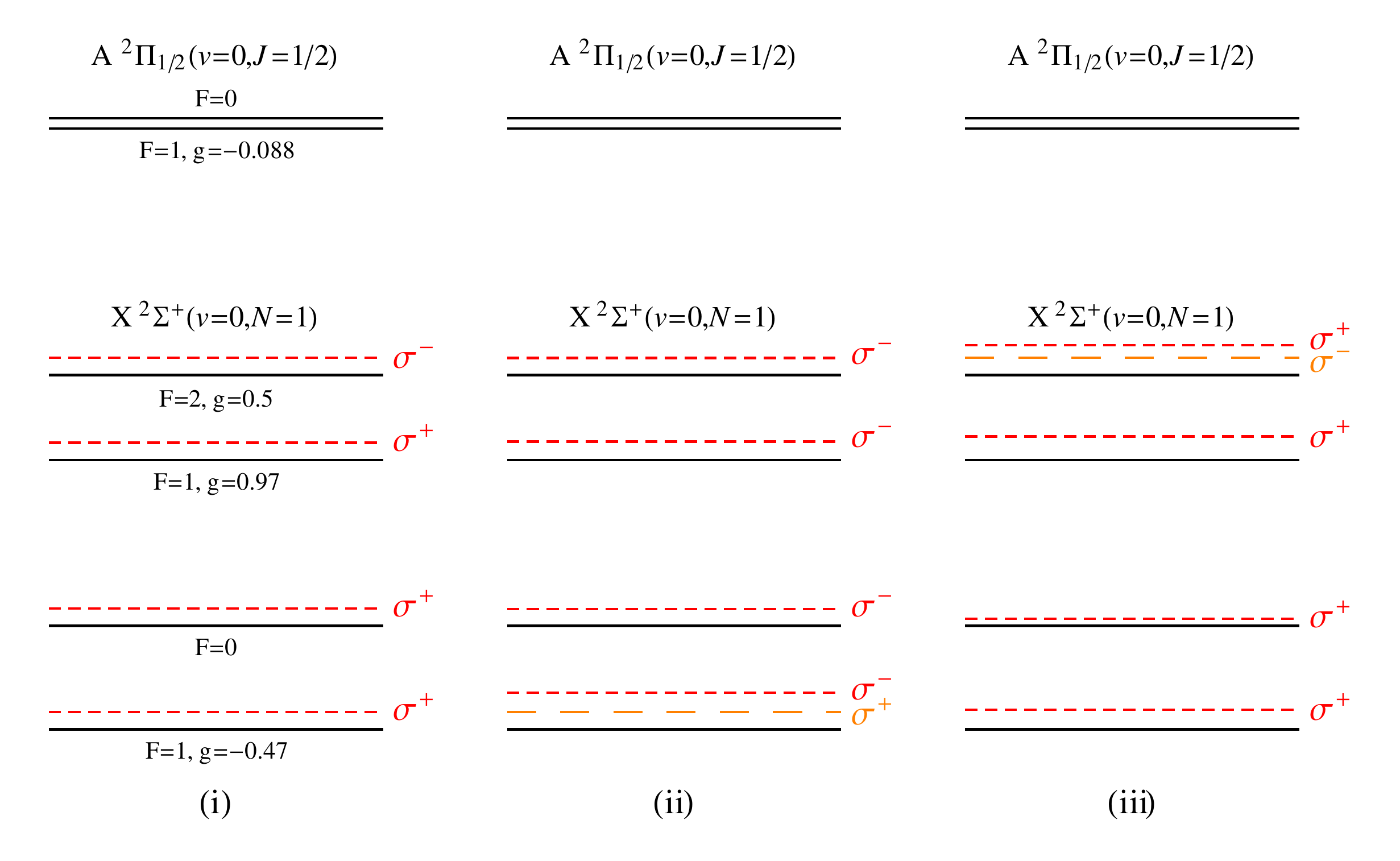}
		\caption{Schemes for laser cooling and magneto-optical trapping of SrF. Solid lines show the relevant energy levels in the X\,$^{2}\Sigma^{+}(v=0, N=1)$ ground state and the A$^{2}\Pi_{1/2}(v'=0,J'=1/2)$ excited state. In (i), they are labelled by their angular momenta, and their g-factors are given. Dashed lines indicate the frequencies and polarizations of laser components that drive transitions between the X and A states (long dash for ${\cal L}_{00}^{\dagger}$ and short dash for  ${\cal L}_{00}$). (i) Ideal polarizations, following table \ref{tab:accelSummary}. The four laser components addressing the 0-0 vibrational transitions are tuned to resonance and then all detuned by $-1.2\Gamma$. The four laser components addressing the 1-0 vibrational transitions are tuned to resonance. (ii) Setup used in \cite{Barry(1)14}. ${\cal L}_{00}^{\dagger}$ is detuned by $-1.2\Gamma$ from the highest frequency transition. The EOM that generates the ${\cal L}_{00}$ components is modulated at 40.4\,MHz and the centre frequency chosen to minimize the quadrature sum of the detunings from the upper 3 levels and then offset by by $-1.2\Gamma$. The EOM that generates the ${\cal L}_{10}$ components is modulated at 42.5\,MHz and the centre frequency chosen to minimize the quadrature sum of the detunings from all 4 levels. (iii) Modification to the scheme in setup (ii) to more closely approximate setup (i). ${\cal L}_{00}^{\dagger}$ is detuned by $-1.2\Gamma$ from the lowest frequency transition. The EOM that generates the ${\cal L}_{00}$ components is modulated at 42.9\,MHz and the centre frequency chosen to minimize the quadrature sum of the detunings from the lower 3 levels and then offset by by $-1.2\Gamma$. The ${\cal L}_{10}$ frequencies are the same as in (ii)}
		\label{fig:SrFSchemes}
\end{figure*}

Let us now focus on the specific case of a MOT for SrF, operating on the X\,$^{2}\Sigma^{+}(v, N=1)$ to A$^{2}\Pi_{1/2}(v'=0,J'=1/2)$ transition, since this has recently been demonstrated \cite{Barry(1)14}. In the experiment, the two lowest vibrational levels ($v=0,1$) are both driven to the same upper level. Decays to higher-lying vibrational levels ($v>1$) are rare and are repumped through different vibrational levels of the A state. We neglect these rare decays which we expect to have very little influence. Each vibrational level of the ground state has 4 levels due to the hyperfine and spin-rotation interactions, while the upper state has 2 hyperfine levels whose splitting is smaller than $\Gamma$. This structure is shown in figure \ref{fig:SrFSchemes}. The simulation takes into account the 4 Zeeman sub-levels of the A state, and the 24 lower sub-levels of X (12 each for $v=0$ and 1). The transition amplitudes between the lower and upper levels are calculated using the known Franck-Condon factors and standard angular momentum algebra (see the appendix of \cite{WallFC}). The upper state g-factor is set to the value given in table \ref{tab:gfacs}. The powers, detunings, and waist sizes of all the laser beams, and the magnetic field gradient, are all set to the values used in \cite{Barry(1)14}.

\begin{figure}[tb]
	\centering
		\includegraphics[width=0.6\textwidth]{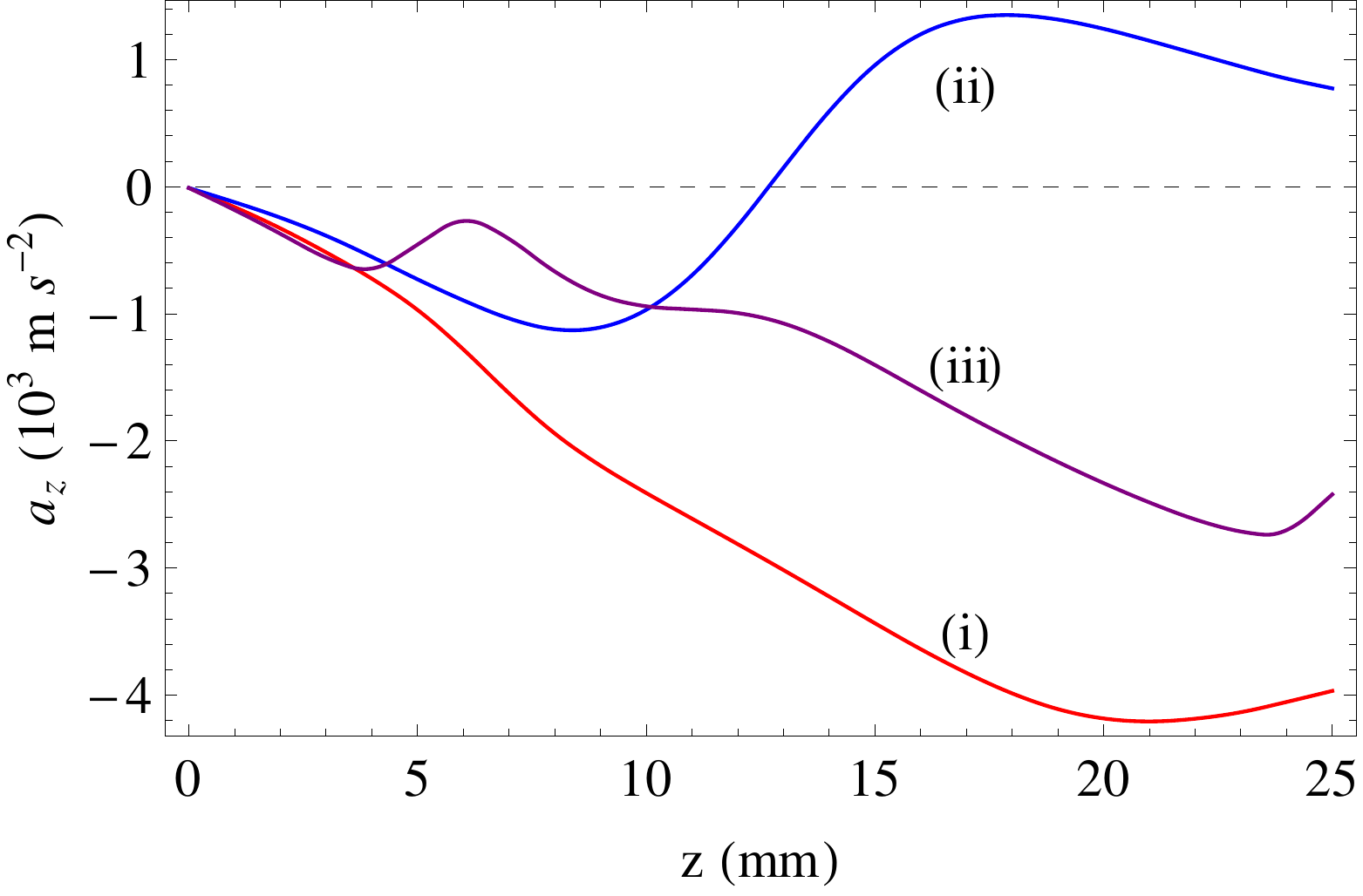}
		\caption{Acceleration versus axial position for the three schemes shown in figure \ref{fig:AccelCurvesSrFSchemes}. Parameters are $\lambda = 663.3$\,nm, $\Gamma=2\pi\times 7$\,MHz, $m=107$\,amu, $A=7.5$\,G/cm, $w=7$\,mm. The power in each of the 4 (or 5) components of the main cooling cycle is 50\,mW, and in each of the 4 components of the vibrational repump is 40\,mW.}
		\label{fig:AccelCurvesSrFSchemes}
\end{figure}

The best choice of polarizations for driving the various transitions is far from obvious in the complicated case where all the transitions share a common upper level. Nevertheless, it seems reasonable to follow the prescriptions for the individual angular momentum cases detailed above and summarized in table \ref{tab:accelSummary}. Since the upper state g-factor is negative, the restoring beams should drive $\Delta M=-1$ transitions for the $F_{l}=2$ to $F_{u}=1$ system, and should drive $\Delta M=+1$ transitions for all the other cases. These choices of polarizations are indicated in figure \ref{fig:SrFSchemes}(i). The acceleration as a function of $z$ for this arrangement is shown by curve (i) in figure \ref{fig:AccelCurvesSrFSchemes}. From the slope of this curve close to the origin we find a trapping frequency of $\omega_{z}=2\pi\times 60$\,Hz. Although the $1/e^{2}$ radius of the beams is only 7\,mm, there is a confining force all the way out to $z=25$\,mm and beyond, showing that the orthogonal beams are not essential for setting up the confining force in this complicated case. Reversing the polarizations of any one of the laser components reduces the trapping frequency, suggesting that this arrangement is indeed the best one. Reversing the polarization of the component that drives the transition from $F_{l}=2$ has a particularly strong effect, reducing $\omega_{z}$ by a factor of 3, suggesting that it is mainly this transition that produces the confinement.

It is difficult in practice to arrange the setup shown in figure \ref{fig:SrFSchemes}(i). The actual scheme used in the experiment \cite{Barry(1)14} is shown in figure \ref{fig:SrFSchemes}(ii). The four frequency components shown in red are identically polarized, and are derived from a single laser (called ${\cal L}_{00}$) using the 1st and 2nd order sidebands of an electro-optic modulator (EOM). They primarily drive the transitions from the $F=2$, $F=0$ and upper $F=1$ levels. An additional laser ${\cal L}_{00}^{\dagger}$, oppositely polarized, drives the transition from the lower $F=1$ level. The transitions from $v=1$ are driven by a third laser (${\cal L}_{10}$), and the components generated by an EOM modulated so that all four components are close to resonance. Simulating this setup, we obtain the acceleration shown by curve (ii) in figure \ref{fig:AccelCurvesSrFSchemes}. For $z<12.5$\,mm, this arrangement produces a confining force, and close to the origin we find a trapping frequency of $\omega_{z}=2\pi\times 51$\,Hz, about twice the value actually found in the experiment. This $\omega_{z}$ is only a little smaller than in case (i), probably because the beam addressing the $F=2$ transition has the same polarization as in (i), and provides most of the confinement. As in the experiment, the MOT disappears if all the polarizations are reversed. It also disappears if the polarization of ${\cal L}_{00}^{\dagger}$ is reversed, or if this laser is turned off, showing the importance of this extra frequency component in this scheme. At larger distances from the centre the force becomes de-confining. This happens close to the $B$ field where the upper $F=1$, $M=1$ level crosses the $F=2, M=-1,-2$ levels, and this seems to be the reason for the change in the sign of the force. Simulating a distribution of molecules in the MOT, the mean scattering rate is found to be $(4.2 \pm 0.1) \times 10^{6}$ photons/s, consistent with the value found in the experiment. The equilibrium temperature is $1.1\pm 0.2$\,mK, which is about half the measured temperature. The damping coefficient in the simulation is found to be $\beta =2030$\,s$^{-1}$, a factor of 15 higher than in the experiment.

Figure \ref{fig:SrFSchemes}(iii) shows an alternative way of using ${\cal L}_{00}$ and ${\cal L}_{00}^{\dagger}$ so that the setup more closely approximates scheme (i). Here, the EOM is modulated at 42.9\,MHz and the centre frequency chosen to minimize the quadrature sum of the detunings from the $F=0$ and $F=1$ components. The ${\cal L}_{00}^{\dagger}$ laser is tuned into resonance with the $F=2$ component. Then, all components are detuned by $-1.2\Gamma$. The acceleration curve found for this setup is shown by curve (iii) in figure \ref{fig:AccelCurvesSrFSchemes}. Its gradient close to the origin is almost identical to case (i) and is an improvement over case (ii). The force is also confining over the whole range of $z$ explored, and this may improve the capture volume of the MOT.

In all three cases considered, the trapping forces are weak relative to a conventional atomic MOT. This is partly because of the small upper state $g$-factor, and partly because 3 out of the 4 transitions being driven have $F_{u} \le F_{l}$ which always give weaker confinement relative to the preferred case of $F_{u} > F_{l}$ (see section \ref{Sec:AngMomCases}). It is likely that the rapid switching method will increase the confining forces considerably.

\section{Summary and conclusions}

This paper explains the essential mechanisms of magneto-optical trapping for atoms and molecules that have complex energy level structures. We have used qualitative pictures to build an understanding, and have given quantitative results based on rate equations. This model cannot capture the effects of coherences between levels which may be important in a real MOT. Of particular relevance is coherent population trapping into dark states, and the mechanisms that destabilize those dark states. We have assumed linear Zeeman shifts, neglecting the mixing of hyperfine states by the magnetic field which is know to be important for closely-spaced hyperfine levels \cite{Nasyrov(1)01}. We have also neglected the interference between the laser beams, which sets up polarization gradients that are likely to be important. Though these effects may change some of the quantitative results obtained here, they are unlikely to change the main conclusions which are based on simple considerations about how the trapping forces arise. We now summarize those conclusions.

The strongest forces are always obtained for $F_{u} > F_{l}$, because only then can the atoms interact continuously with the restoring beam without being pumped into a state that is dark to that beam. We find that when $F_{u} \ge F_{l}$ the laser polarizations should be chosen as in a normal atomic MOT, but when $F_{u} < F_{l}$ they have to be reversed. The polarizations also have to be reversed if the light is blue-detuned, or if the upper-state g-factor is negative, but not if the lower state g-factor is negative. These polarization rules explain some previously unexplained MOT configurations that appear in the literature, e.g. \cite{Flemming(1)97}. When the upper state has no Zeeman splitting, the forces average to zero. Similarly, when $g_{u} \ll g_{l}$ the trapping forces are very weak.

In cases where the confining forces are weak, either because of a small upper state g-factor, or because $F_{u} \le F_{l}$, a stronger force can be generated by rapidly and synchronously reversing the field gradient and laser polarizations. The switch must be fast enough that the change is non-adiabatic, and the strongest force will be obtained when the switching period is comparable to the optical pumping time. Figure \ref{fig:accelCurvesRapidSwitching} shows the forces obtained from simulations where this rapid switching method is applied.

These conclusions are particularly relevant for magneto-optical trapping of molecules where the laser cooling uses a $^{2}\Sigma - ^{2}\Pi_{1/2}$ transition. A pure $^{2}\Pi_{1/2}$ state has a very small g-factor, too small to make a dc MOT. We have estimated the actual g-factor for several relevant molecules, and find that it can be far larger than that because of mixing with a nearby $^{2}\Sigma$ state. For SrF, the upper state g-factor is large enough to produce a weakly-confining MOT, as observed experimentally \cite{Barry(1)14}. Using the rate equation model, we find a trapping frequency, scattering rate, and equilibrium temperature all within a factor of 2 of those found in the experiment. However, the damping coefficient found in this simulation is an order of magnitude higher than measured experimentally. We suggest that the polarizations used in \cite{Barry(1)14} are not the optimum ones, and suggest a simple change to the setup that will match the optimum arrangement more closely. This alternative setup should give a higher trapping frequency and larger capture volume.

\ack
I am grateful to Ed Hinds, Ben Sauer, Jony Hudson and Tim Steimle for helpful advice and discussions.

\section*{References}
\bibliographystyle{iopart-num}

\end{document}